# Electromagnetic energy, momentum and forces in a dielectric medium with losses

Yurii A. Spirichev

The State Atomic Energy Corporation ROSATOM, "Research and Design Institute of Radio-Electronic Engineering" - branch of Federal Scientific-Production Center "Production Association "Start" named after Michael V.Protsenko",

Zarechny, Penza region, Russia

E-mail: yurii.spirichev@mail.ru

**Abstract**

From the energy-momentum tensors of the electromagnetic field and the mechanical energy-momentum, the equations of energy conservation and balance of electromagnetic and mechanical forces are obtained. The equation for the Abraham force in a dielectric medium with losses is obtained

The contents

1. Introduction
2. Energy-momentum tensors
3. Equations of conservation of energy and momentum
4. Electromagnetic forces
5. Conclusion
   References

## 1. Introduction

The problem of electromagnetic momentum and electromagnetic forces arising as a result of the interaction of electromagnetic radiation with a dielectric medium is the subject of numerous studies. Some recent papers on the theme are listed in Refs. [1–15] et al. In these articles, the forms of the electromagnetic momentum in the medium, the shapes of the electromagnetic energy-momentum tensors (EMT), the Abraham force and other electromagnetic forces are considered. In article [6] from tensors of electromagnetic field and electromagnetic induction the EMT of interaction of an electromagnetic field with a dielectric medium and the equations of conservation of energy and momentum in a dielectric medium without losses are obtained. Equations of conservation of the electromagnetic momentum are simultaneously the equations of the balance of electromagnetic forces of interaction of electromagnetic radiation with the medium. Here two cases are possible.

In the first case, when light passes through a transparent medium, there are no losses, i.e. absorption and scattering of light. Then the beam electromagnetic energy remains unchanged before entering the medium and after exiting it, and the divergences of the EMT in the medium and outside the medium equal zero, since there is no electromagnetic energy sink. In this case, the equations of conservation of only electromagnetic energy and momentum follow from EMT. If there are no losses in the medium, then there is only an exchange of reactive energy between the wave and the medium. In this case, the forces arising in the medium are reactive forces and they do not perform work in it. This case is considered in [6].

In the second case, when light passes through the medium, there are losses. Then part of the electromagnetic radiation energy remains in the medium. In this case, the divergence of the electromagnetic energy-momentum tensor is not zero, since there is a sink of electromagnetic energy. In this case, mechanical forces arise that perform work on deformation, heating, or motion of the medium. This work consumes electromagnetic energy of radiation. In this case, the equations of conservation of electromagnetic energy and momentum must be supplemented by the mechanical energy and momentum of the medium, which are described by the mechanical energy-momentum tensor. In this case, the divergence of the mechanical energy tensor is the source, and the divergence of the electromagnetic energy-momentum tensor is a sink. Then the process of transferring of electromagnetic energy to the medium can be written in the form of equations $\partial^\mu T_{\mu\nu}^E = \partial^\mu T_{\mu\nu}^M$ and $\partial^\nu T_{\mu\nu}^E = \partial^\nu T_{\mu\nu}^M$, where $T_{\mu\nu}^E$ is the electromagnetic energy-momentum tensor, $T_{\mu\nu}^M$ is the mechanical energy-momentum tensor (divergences for different indices have differences, since the tensor is asymmetric). In this case, the interaction of light with the medium is reactive and active (or complex).

The purpose of this paper is to consider electromagnetic energy, momentum, and electromagnetic forces in a dielectric medium with losses.

**2. Energy-momentum tensors**

We can write the canonical EMT in its general form [16]

$$T_{\mu\nu} = \begin{bmatrix} W & i\frac{1}{c}\mathbf{S} \\ ic\cdot\mathbf{g} & t_{ik} \end{bmatrix} \quad (\mu, \nu = 0, 1, 2, 3;\ i, k = 1, 2, 3) \quad (1)$$

где  W – The Energy density;

**S** – The Energy flux density (Poynting vector);

**g** – The Momentum density;

$t_{ik}$ – The tensor of the momentum flux density (the stress tensor)

The components of the EMT obtained in [6] have the form:

$W = \mathbf{E}\cdot\mathbf{D}$  $\quad\quad\quad \mathbf{S} = \mathbf{E}\times\mathbf{H}$

$\mathbf{g} = \mathbf{D}\times\mathbf{B}$  $\quad\quad\quad t_{ik} = E_i D_k + B_i H_k - 3\delta_{ik}(\mathbf{B}\cdot\mathbf{H})$.

After their substitution into the canonical tensor (1), the new EMT has the form:

$$T_{\mu\nu} = \begin{bmatrix} \mathbf{E}\cdot\mathbf{D} & i\frac{1}{c}(\mathbf{E}\times\mathbf{H}) \\ ic\cdot(\mathbf{D}\times\mathbf{B}) & E_i D_k + B_i H_k - 3\delta_{ik}(\mathbf{B}\cdot\mathbf{H}) \end{bmatrix} \quad (2)$$

It differs from Minkowski's EMT by diagonal terms.: Its linear invariant $I_1$ is the the Lagrange density function for electromagnetic field in the medium, and its linear invariant for the microfield or vacuum $I_2$ is a quadratic invariant of the antisymmetric electromagnetic field tensor:

$$I_1 = 2(\mathbf{E}\cdot\mathbf{D} - \mathbf{B}\cdot\mathbf{H}) \qquad I_2 = 2(\varepsilon_0 \mathbf{E}^2 - \mathbf{B}^2/\mu_0)$$

The isotropic dielectric medium is described by the canonical material equations $\mathbf{D} = \varepsilon\cdot\varepsilon_0\cdot\mathbf{E}$ and $\mathbf{H} = \mathbf{B}/\mu\cdot\mu_0$, where ε and μ, respectively, the relative dielectric and magnetic permeability's of the medium.

$$T_{\mu\nu} = \begin{bmatrix} \varepsilon\varepsilon_0 \mathbf{E}^2 & i\cdot(\mathbf{E}\times\mathbf{B})/c\cdot\mu\cdot\mu_0 \\ i\cdot c\cdot\varepsilon\varepsilon_0\cdot(\mathbf{E}\times\mathbf{B}) & \varepsilon\varepsilon_0 \mathbf{E}^2 - 2\cdot\mathbf{B}^2/\mu\cdot\mu_0 \end{bmatrix} \quad (3)$$

For vacuum $\varepsilon = \mu = 1$. Then the EMT (2) for vacuum can be written in the form:

$$T_{\mu\nu}^V = \begin{bmatrix} \varepsilon_0\cdot\mathbf{E}^2 & i\cdot(\mathbf{E}\times\mathbf{B})/c\cdot\mu_0 \\ i\cdot c\cdot\varepsilon_0\cdot(\mathbf{E}\times\mathbf{B}) & \varepsilon_0\cdot\mathbf{E}^2 - 2\cdot\mathbf{B}^2/\mu_0 \end{bmatrix} \quad (4)$$

The mechanical energy-momentum tensor is obtained as the tensor product of the vector of the four-dimensional velocity of the medium motion $\mathbf{V}_\mu(c, i\cdot\mathbf{V})$ to the four-dimensional density vector of the mechanical momentum $\mathbf{P}_\nu(c\cdot m, i\cdot\mathbf{p})$, where $\mathbf{m}$ is the mass density, $\mathbf{p}$ is the density of the mechanical medium motion. Since the scalar products of the mixed velocity components are zero, we obtain a symmetric mechanical energy-momentum tensor in the form:

$$T_{\mu\nu}^M = \mathbf{V}_\mu \otimes \mathbf{P}_\nu = \begin{bmatrix} W & \frac{1}{c}i\cdot\mathbf{S} \\ i\cdot c\cdot\mathbf{g} & t_{ik} \end{bmatrix} = \begin{pmatrix} m\cdot c^2 & i\cdot c\cdot p_x & i\cdot c\cdot p_y & i\cdot c\cdot p_z \\ i\cdot c\cdot p_x & -V_x\cdot p_x & 0 & 0 \\ i\cdot c\cdot p_y & 0 & -V_y\cdot p_y & 0 \\ i\cdot c\cdot p_z & 0 & 0 & -V_z\cdot p_z \end{pmatrix} \quad (5)$$

Its components have the form:

$$W = m\cdot c^2 \qquad \mathbf{S} = \mathbf{p}\cdot c^2 \qquad \mathbf{g} = \mathbf{p} \qquad t_{ik} = t_{ii} = -\mathbf{v}\cdot\mathbf{p}$$

Its linear invariant has the form:

$$I = m\cdot c^2 - \mathbf{v}\cdot\mathbf{p}$$

After the transition from mass density to mass this invariant corresponds to the canonical expression known in mechanics as a canonical expression of the relativistic Lagrange function [16] for a free particle taken with the opposite sign.

**3. Equations of conservation of energy and momentum in a medium with losses**

The equations for the conservation of energy and momentum in a medium with losses follow from the EMT (2) and (5) in the form of their four-dimensional divergences $\partial^\mu T_{\mu\nu}^E = \partial^\mu T_{\mu\nu}^M$ and $\partial^\nu T^E_{\mu\nu} = \partial^\nu T^M_{\mu\nu}$. In the general case, the EMT (2) is asymmetric and for each of its indices one can write down two groups of equations:

$$\frac{1}{c}\partial_t(\mathbf{E}\cdot\mathbf{D}) + c\cdot\nabla\cdot(\mathbf{D}\times\mathbf{B}) = \partial_t m + \nabla\cdot\mathbf{p} \qquad (6)$$

$$\frac{1}{c^2}\partial_t(\mathbf{E}\times\mathbf{H}) - \partial_i(E_i D_k + B_i H_k - 3\delta_{ik}(\mathbf{B}\cdot\mathbf{H})) = \partial_t\mathbf{p} + \nabla(\mathbf{p}\cdot\mathbf{V}) \qquad (7)$$

$$\frac{1}{c}\partial_t(\mathbf{E}\cdot\mathbf{D}) + \nabla\cdot(\mathbf{E}\times\mathbf{H})/c = \partial_t m + \nabla\cdot\mathbf{p} \qquad (8)$$

$$\partial_t(\mathbf{D}\times\mathbf{B}) - \partial_k(E_i D_k + B_i H_k - 3\delta_{ik}(\mathbf{B}\cdot\mathbf{H})) = \partial_t\mathbf{p} + \nabla(\mathbf{p}\cdot\mathbf{V}) \qquad (9)$$

Eq. (6) and (8) are equations of the balance of electromagnetic and mechanical energy. Eq. (7) and (9) are equations of the balance of electromagnetic and mechanical momentum. In Eq. (6) and (8), (7) and (9), the right-hand sides of the equations are equal. This is due to the symmetry of the mechanical energy-momentum tensor. From the equations of conservation of the energy density (6) and (8) follows the equation:

$$c\cdot\nabla\cdot(\mathbf{D}\times\mathbf{B}) = \nabla\cdot(\mathbf{E}\times\mathbf{H})/c \quad\text{or}\quad \nabla\cdot(\mathbf{D}\times\mathbf{B}) = \nabla\cdot(\mathbf{E}\times\mathbf{H})/c^2 \qquad (10)$$

The left side of this equation is the divergence of the momentum density in the Minkowski's form, and the right-hand side is the divergence of the momentum density in the Abraham's form.

From Eq. (7) and (9), taking into account the equality of their right-hand sides follows the equation:

$$\partial_t(\mathbf{D}\times\mathbf{B}) - \partial_k(E_i D_k + B_i H_k - 3\delta_{ik}(\mathbf{B}\cdot\mathbf{H})) = \frac{1}{c^2}\partial_t(\mathbf{E}\times\mathbf{H}) - \partial_i(E_i D_k + B_i H_k - 3\delta_{ik}(\mathbf{B}\cdot\mathbf{H})) \qquad (11)$$

**4. Electromagnetic forces in a dielectric medium with losses**

Electromagnetic forces, more precisely, the density of electromagnetic forces in a continuous dielectric medium with losses are determined in the form of derivatives of the density of the electromagnetic momentum with respect to time $\partial_t\mathbf{g}$. Then Eq. (7) and (9) can be regarded as equations of the balance of electromagnetic and mechanical forces in the medium. Eq. (7) for the momentum density in the Abraham's form can be written in the form of a balance of electromagnetic and mechanical forces:

$$\partial_t\mathbf{g}^A - \partial_i(E_i D_k + B_i H_k - 3\delta_{ik}(\mathbf{B}\cdot\mathbf{H})) - \partial_t\mathbf{p} - \nabla(\mathbf{p}\cdot\mathbf{V}) = 0 \qquad (12)$$

Eq. (9) for the momentum density in the Minkowski's form can be written in the form of a balance of electromagnetic and mechanical forces:

$$\partial_t\mathbf{g}^M - \partial_k(E_i D_k + B_i H_k - 3\delta_{ik}(\mathbf{B}\cdot\mathbf{H})) - \partial_t\mathbf{p} - \nabla(\mathbf{p}\cdot\mathbf{V}) = 0 \qquad (13)$$

Balance force equations are distinguished by the second terms. Let us find this difference in the form of the difference between Eq. (13) and (12):

$$\partial_t \mathbf{g}^M - \partial_t \mathbf{g}^A = \partial_k (E_i D_k + B_i H_k - 3\delta_{ik}(\mathbf{B} \cdot \mathbf{H})) - \partial_i (E_i D_k + B_i H_k - 3\delta_{ik}(\mathbf{B} \cdot \mathbf{H})) =$$
$$= \nabla \times (\mathbf{E} \times \mathbf{D} + \mathbf{B} \times \mathbf{H})$$

The left side of this balance force equation is the Abraham force. Final equation for the Abraham force in a dielectric medium with losses has the form:

$$\mathbf{F}_A = \nabla \times (\mathbf{E} \times \mathbf{D} + \mathbf{B} \times \mathbf{H}) \tag{14}$$

This equation does not differ from the equation of the Abraham force for a lossless medium [6]. From Eq. (12) and (13) it follows that in the presence of losses, electromagnetic forces because mechanical forces in the medium that perform work on motion, deformation and heating of the medium. Eq. (14) for the Abraham force shows that it is a vortex reactive force that does not depend on the losses in the medium.

If the medium is described by canonical material equations of the form $\mathbf{D} = \varepsilon \cdot \varepsilon_0 \cdot \mathbf{E}$ and $\mathbf{H} = \mathbf{B}/\mu \cdot \mu_0$ and $\varepsilon$ and $\mu$ are constants or scalar functions, then the vectors $\mathbf{D}$ and $\mathbf{E}$, $\mathbf{H}$ and $\mathbf{B}$ are collinear and the right-hand side of the Eq. (14) with vector products of these vectors is zero. Then the Abraham force is zero.

If the medium is described by the canonical material equations, then the non-diagonal components of the stress tensor $t_{ik}$ are zero, and the electromagnetic forces influencing the medium are determined only by its diagonal components. Then $\partial_t \mathbf{g}^M = \partial_t \mathbf{g}^A = \partial_t \mathbf{g}$ the Eq. (8) and (10) of the forces balance in the medium can be written in the form of a single equation:

$$\partial_t \mathbf{g} - \nabla(\varepsilon \cdot \varepsilon_0 \cdot \mathbf{E}^2 - 2\mu \cdot \mu_0 \cdot \mathbf{H}^2) = \partial_t \mathbf{p} + \nabla(\mathbf{p} \cdot \mathbf{V}) \tag{15}$$

Applying vector identities to this equation, we obtain electromagnetic and mechanical forces in the expanded form:

$$\partial_t \mathbf{g} - (\nabla \varepsilon) \cdot \varepsilon_0 \cdot \mathbf{E}^2 - \varepsilon \cdot \varepsilon_0 \cdot (\mathbf{E} \cdot \nabla)\mathbf{E} - \varepsilon \cdot \varepsilon_0 \cdot \mathbf{E} \times \nabla \times \mathbf{E} - 2(\nabla \mu) \cdot \mu_0 \cdot \mathbf{H}^2 - 2\mu \cdot \mu_0 \cdot (\mathbf{H} \cdot \nabla)\mathbf{H} - 2\mu \cdot \mu_0 \cdot \mathbf{H} \times \nabla \times \mathbf{H} =$$
$$= \partial_t \mathbf{p} + (\mathbf{p} \cdot \nabla)\mathbf{V} + (\mathbf{V} \cdot \nabla)\mathbf{p} + \mathbf{p} \times \nabla \times \mathbf{V} + \mathbf{V} \times \nabla \times \mathbf{p}$$

This equation describes the total balance of electromagnetic and mechanical forces in a dielectric medium with losses. In its left side there are electromagnetic forces, and on the right side there are mechanical forces. For constant $\varepsilon$ and $\mu$ this equation is simplified:

$$\partial_t \mathbf{g} - \varepsilon \cdot \varepsilon_0 \cdot (\mathbf{E} \cdot \nabla)\mathbf{E} - \varepsilon \cdot \varepsilon_0 \cdot \mathbf{E} \times \nabla \times \mathbf{E} - 2\mu \cdot \mu_0 \cdot (\mathbf{H} \cdot \nabla)\mathbf{H} - 2\mu \cdot \mu_0 \cdot \mathbf{H} \times \nabla \times \mathbf{H} =$$
$$= \partial_t \mathbf{p} + (\mathbf{p} \cdot \nabla)\mathbf{V} + (\mathbf{V} \cdot \nabla)\mathbf{p} + \mathbf{p} \times \nabla \times \mathbf{V} + \mathbf{V} \times \nabla \times \mathbf{p}$$

Comparing individual electromagnetic and mechanical forces in this equation one can see their complete correspondence to each other.

**7 Conclusion**

The description of energy, electromagnetic momentum and electromagnetic forces in a dielectric medium with losses are obtained. It is shown that the Abraham force is a reactive force and does not depend on losses in the dielectric medium. A complete detailed balance equation for all electromagnetic and mechanical forces in a dielectric medium with losses is obtained.